\documentclass[5p,authoryear]{elsarticle}

\usepackage{amsmath,amssymb,mathtools,siunitx,etoolbox} 
\usepackage{graphicx,enumitem,xcolor,booktabs,tikz,chemplants,natbib}
\usepackage[scale=0.8]{noto-mono}
\usepackage[cachedir=minted-cache]{minted}
\usepackage[colorlinks,allcolors=blue]{hyperref}
\usepackage[capitalize]{cleveref}

\usetikzlibrary{arrows, arrows.meta}
\journal{arXiv}
\setlist[itemize]{noitemsep}
\setminted[julia]{%
    fontsize=\small,
    bgcolor=gray!10,
}
\setminted[julia-repl]{%
    fontsize=\small,
}
\newcommand*{\spacerepl}{\vspace{-1.5em}}
\begin{document}

\begin{frontmatter}

\title{ModelPredictiveControl.jl: advanced\\process control made easy in Julia}

\author[jumineandvendel]{Francis Gagnon}\corref{corresponding}\ead{franckgaga2@gmail.com}
\author[jumine]{Alex Thivierge}
\author[loop]{André Desbiens}
\author[juliahub]{Fredrik Bagge Carlson}

\address[jumineandvendel]{Work mainly performed while at: Jumine, Quebec City, Canada. Currently at: Vendel, Quebec City, Canada}
\address[jumine]{Jumine, Quebec City, Canada}
\address[loop]{Process Observation and Optimization Laboratory (LOOP), Université Laval, Quebec City, Canada}
\address[juliahub]{JuliaHub, Lund, Sweden}

\begin{abstract}
Proprietary closed-source software is still the norm in advanced process control. Transparency and reproducibility are key aspects of scientific research. Free and open-source toolkit can contribute to the development, sharing and advancement of  new and efficient control approaches, and the industrial sector will certainly benefit from them. This paper presents ModelPredictiveControl.jl, an open-source software package for designing model predictive controllers in the Julia programming language. It is designed to be easy to use and modular, while providing advanced features like nonlinear control and moving horizon estimation. It relies on powerful control system, mathematical optimization and automatic differentiation frameworks to simplify the construction and testing of state estimators and predictive controllers. It also integrates with the standard plotting library to quickly visualize closed-loop data. The paper presents the main functionalities and illustrates them with two case studies in simulation. The first example is a continuously stirred tank reactor described by linear dynamics. The second one implements a nonlinear, an economic, and a successive linearization model predictive controllers for an inverted pendulum. The solving times are benchmarked against equivalent implementations in MATLAB to show the efficiency of the package.
\end{abstract}
    
\cortext[corresponding]{Corresponding author.}


\end{frontmatter}

% !TeX encoding = UTF-8
% !TeX spellcheck = en_US
\section{Introduction}

The process control community, both in the academic and industrial sectors, has largely relied on MATLAB toolboxes and commercial solutions for designing, simulating and implementing closed-loop systems \citep{optimMatlab}. The ecosystem developed by MathWorks\texttrademark\ is rich, mature, cohesive, and well documented, but their licensing policy can be unaffordable for smaller organizations. Moreover, because it is a proprietary software, the source code of many functions is not available. This is an issue for scientific research, where reproducibility and transparency is a key aspect. The dependency on a single vendor also raises concerns about the long-term support of third-party control software products. Lastly, like any interpreted languages, its performance can be suboptimal for computationally intensive tasks, especially for time-critical applications like real time optimization and model predictive control \citep{matlabPythonJulia, juliaML}. Code generation mitigates this issue, but it creates other problems like additional licensing fees, the difficulty to debug the generated code, and the restriction to a subset of the language. There are multiple efforts for open source toolkits developed in Python or C \citep{doMPC, gekko, acados}, but the languages are not particularly \emph{tailored} for numerical computing, particularly when it comes to linear algebra. The interface also frequently limits the user when implementing the model or custom objectives/constraints. For instance, tools based on CasADi \citep{casadi} force the user to code the dynamics into something supported by the framework.

Julia is relatively new programming language specialized for scientific and numerical computing. The just-ahead-of-time compiler can reach performance comparable to C and Fortran, while exhibiting a modern and expressive syntax like MATLAB and Python \citep{juliaPaper}. The built-in read-eval-print-loop (REPL) allows to interactively test code and inspect variables, mimicking the development workflow of an interpreted language. It also includes a package manager with a general registry, which makes it easy to share, install or update code. Moreover, the environment is free and open-source and it can be used for commercial purposes without any licensing fees. The ecosystem is still young, but packages for control systems (\texttt{ControlSystems.jl}), system identification (\texttt{ControlSystemIdentification.jl}), optimization (\texttt{JuMP.jl}) and automatic differentiation through high-level abstractions (\texttt{DifferentiationInterfaces.jl}) are already available \citep{controlsystems, jump, differentiationinterface}. At the start of this work, there was no free and open-source MPC package available in the official registry of Julia, which is the gap this package fills. A library specialized for linear MPC (\texttt{LinearMPC.jl}) has recently become available, and it is now leveraged to generate lightweight C code for embedded platforms.

This paper presents a general-purpose MPC package for Julia. It aims to provide a simple, clear and modular framework to quickly design and test predictive controllers in Julia, while preserving the flexibility for advanced real-time optimization. It does not particularly strive for original methodological scientific contributions or bleeding edge technologies like robust MPC, but rather to improve the accessibility to advanced process control, both for the academic and industrial community. Some sectors like the petrochemical industry have been using MPC for decades, but other industries like mineral and pharmaceutical processing are lagging behind in terms of online quality control and optimization \citep{gapMPC, gapMPC2}. These tools can significantly reduce the waste in raw material and energy consumption. This is highly needed in the context of climate changes and the increasing scarcity of resources.

It focuses on modern MPCs that rely on a closed-loop state estimator for the feedback. The package currently implements all the classical Kalman-type filters. It also incorporates an internal model control structure, as a more traditional approach. In addition, the analog of predictive control but for state estimation, the moving horizon estimator (MHE), is also available to solve constrained estimation problems. The user can also disable the built-in observer and provide its own feedback strategy.

Linear plant models are automatically augmented with an appropriate representation of the unmeasured disturbances based on observability (customizable). Additionnaly, C code generation is available for linear controllers via \texttt{LinearMPC.jl}, which can be used for real-time applications on embedded platforms. For the nonlinear case, the user chooses to add integrated white noise on the input or output for each channel. The \texttt{JuMP.jl} interface allows changing the solver quickly among  many open-source and commercial optimization software (local and global). As generic Julia functions can be differentiated using automatic differentiation, the gradient, Jacobian and Hessian of the objective/constraint function for nonlinear MPCs are computed automatically within machine precision. The \texttt{DifferentiationInterfaces.jl} dependency allows to easily switch between different automatic differentiation or finite difference tools, while efficiently handling the possible sparsity of the problem. This also applies for the extended Kalman Filter, as well as adaptive laws based on successive linearizations of a nonlinear plant model. 

By leveraging the multiple dispatch paradigm of Julia, nonlinear controllers based on linear models (e.g., economic MPC) evaluate the predictions with matrix algebra instead of a \texttt{for} loop, which is generally more efficient especially for the constraint handling. More precisely, when a function of the package is called, the exact implementation that will be executed (a ``method'') depends on both the types of the controller and the plant model, to select the most appropriate block of code. Both soft and hard constraints on inputs, input increments, outputs, and terminal states are supported. The MHE state and noise estimates also support constraint relaxation.

The next section presents the structure of the package. The following one present two case studies to illustrate the syntax, a linear and a nonlinear controller design, and benchmark the performances against MATLAB.
% !TeX encoding = UTF-8
% !TeX spellcheck = en_US

\section{Methodology}

\begin{figure}[ht]
    \centering
    \includegraphics[trim=0 15 0 15, width=0.95\columnwidth]{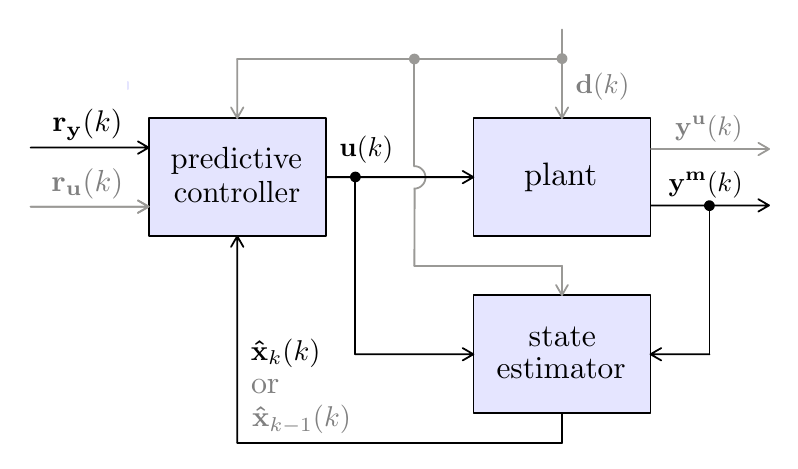}
    \caption{Block diagram of the controllers in this package.}
    \label{fig:control_topology}
\end{figure}

The functions of this package dispatch over three abstract types, described hereafter. The
\cref{fig:control_topology} illustrates the main signals and the interactions between the different components of the controllers. The gray parts are features that are less commonly used, thus disabled by default. The following sections introduce the various mathematical symbols depicted in the figure.

\paragraph{\textnormal{\texttt{SimModel}}} (2 subtypes) -- Discrete state-space models of the plant, including linear and nonlinear representations. Constructors automatically discretize continuous-time linear systems. Instances of \texttt{SimModel} serves as a wrapper to construct \texttt{StateEstimator} or \texttt{PredictiveController} objects, and also as plant simulators to test the designs.

\paragraph{\textnormal{\texttt{StateEstimator}}} (8 subtypes) -- Open loop and closed-loop state observers, both for deterministic or stochastic systems. They produce the full state feedback for the \texttt{PredictiveController}.

\paragraph{\textnormal{\texttt{PredictiveController}}} (3 subtypes)  -- Linear and nonlinear MPC are available. An explicit controller is also possible for linear models without constraint.

\subsection{Plant Models}

Plant models are subtypes of \texttt{SimModel} abstract type. Operating points on the model inputs, outputs and measured disturbances are explicitly defined by the user. Scaling the inputs and outputs can sometimes ease the optimization. It is not supported yet since it is realizable at the user level, but it is planned for future releases. There are currently two concrete subtypes in the package, introduced in the following two sections. 

\paragraph{\textnormal{\texttt{LinModel}}}
Linear state-space representations of the plant. Continuous-time models are discretized using zero-order hold for the manipulated inputs, and Tustin's approximation, for the measured disturbances (sampled continuous signals, usually). This leads to equations of the form:
\begin{subequations}
\begin{align}
    \mathbf{x}(k+1) &= \mathbf{A x}(k) + \mathbf{B_u u}(k) + \mathbf{B_d d}(k) \\
    \mathbf{y}(k)   &= \mathbf{C x}(k) + \mathbf{D_d d}(k)
\end{align}
\end{subequations}
in which the vectors $\mathbf{u}$, $\mathbf{d}$, $\mathbf{x}$ and $\mathbf{y}$ are the manipulated input, measured disturbance, state and output of the model, respectively. Objects are constructed with \texttt{ss} or \texttt{tf} functions from \texttt{ControlSystems.jl}, or by providing the fives state-space matrices directly. All the \texttt{StateEstimator} and \texttt{PredictiveController} objects based on a \texttt{LinModel} representation allow online model modifications of the state-space matrices (see the \cref{sec:successive_linearization} for an example with successive linearizations).

\paragraph{\textnormal{\texttt{NonLinModel}}}
Nonlinear state-space system of the plant described by ordinary differential equations. Differential and algebraic equation systems are not supported yet, but they are planned. A built-in 4th order Runge-Kutta solver with optional supersampling discretizes continuous-time dynamics by default, leading to the following system of equations:
\begin{subequations}
\begin{align}
    \mathbf{x}(k+1) &= \mathbf{f}\big(\mathbf{x}(k), \mathbf{u}(k), \mathbf{d}(k), \mathbf{p} \big) \\
    \mathbf{y}(k)   &= \mathbf{h}\big( \mathbf{x}(k), \mathbf{d}(k), \mathbf{p} \big)
\end{align}
\end{subequations}
with $\mathbf{p}$ encapsulating all the model parameters. It is worth mentioning that the state update $\mathbf{f}$ and output $\mathbf{h}$ functions must be in pure Julia to design nonlinear MPCs with an automatic differentiation backend, like the default \texttt{ForwardDiff.jl} backend \citep{forwardDiff}. The finite difference backends do not have this requirement.

\subsection{State Estimators}

The estimators of the package focus on control applications, that is, relying on the estimates to compute a full state feedback. They all incorporate some kind of integral action by default if feasible, since it is generally desired to eliminate the steady-state error with closed-loop control (offset-free tracking). \cref{sec:case_studies} gives details on that matter.

They are all implemented in the current form (a.k.a. as filter form) by default to improve accuracy and robustness, that is, they all estimates at each discrete time $k$ the states of the current period $\mathbf{\hat{x}}_k(k)$ relying on the newest measurements, also denoted $\mathbf{\hat{x}}(k|k)$. The predictor form is also available however (a.k.a. delayed form). This allow moving the estimator computations after solving the MPC problem, for when the estimations are expensive like in the MHE.

There are seven \texttt{StateEstimator} concrete types available at the time of writing, all supporting measured $\mathbf{y^m}$ and unmeasured $\mathbf{y^u}$ model outputs. The following list presents them.

\paragraph{\textnormal{\texttt{SteadyKalmanFilter}}}
Steady-state form of the Kalman filter, a.k.a asymptotic form. The solution to the algebraic Riccati equation pre-compute the Kalman gain \citep{simon}. This is the default state estimator for controllers based on \texttt{LinModel} objects.

\paragraph{\textnormal{\texttt{KalmanFilter}}}
Time-varying formulation of the Kalman filter. It can evaluate the estimation error covariance in real time or be applied when the plant model is not constant like in adaptive control.

\paragraph{\textnormal{\texttt{Luenberger}}}
Deterministic state observer based on eigenvalue placement in closed-loop. It pre-computes the observer gain with \texttt{place} function from \texttt{ControlSystems.jl}, that implements the method of \citet{placePoles}.

\paragraph{\textnormal{\texttt{UnscentedKalmanFilter}}}
Kalman filter for nonlinear systems relying on the generalized unscented transform \citep{simon}. It propagates the mean and covariance of the noise by approximating the state probability distribution instead of linearizing the plant model like in the \texttt{ExtendedKalmanFilter}. This is the default state estimator for controllers based on \texttt{NonLinModel} objects.

\paragraph{\textnormal{\texttt{ExtendedKalmanFilter}}}
An extension of \texttt{KalmanFilter} for nonlinear models. The Jacobians of the nonlinear state-space functions approximate the propagation of the noise. These matrices are automatically computed by forward mode automatic differentiation.

\paragraph{\textnormal{\texttt{MovingHorizonEstimator}}}
Also known as receding horizon estimation. It minimizes at each discrete time $k$ the following cost function over an estimation window of $N_k = \min(k+1, H_e)$ steps, where $H_e$ is the estimation horizon:
\begin{equation}\label{eq:J_MHE}
    J_{\mathit{MHE}} = \bar{\mathbf{x}}^\intercal \bar{\mathbf{P}}^{-1} \bar{\mathbf{x}} 
    + \mathbf{\hat{W}}^\intercal \mathbf{\hat{Q}}_{N_k}^{-1} \mathbf{\hat{W}}  
    + \mathbf{\hat{V}}^\intercal \mathbf{\hat{R}}_{N_k}^{-1} \mathbf{\hat{V}}
    + C \varepsilon^2
\end{equation}
The state estimate at arrival $\mathbf{\hat{x}}_k(k-N_k+p)$, the process noise estimates over the window $\mathbf{\hat{W}}$ and the slack variable $\varepsilon$ for constraint relaxation are the decision variables. The process and disturbance models compute the sensor noise estimates $\mathbf{\hat{V}}$. The estimation error $\bar{\mathbf{x}} = \mathbf{\hat{x}}_{k-N_k}(k-N_k+p) - \mathbf{\hat{x}}_{k}(k-N_k+p)$ and its covariance $\bar{\mathbf{P}} = \mathbf{\hat{P}}_{k-N_k}(k-N_k+p)$ evaluate the arrival costs. Assigning $p=0$ leads to an estimator in the current formulation (default), $p=1$ for the predictor form. The block diagonal matrices $\mathbf{\hat{Q}}_{N_k}$ and $\mathbf{\hat{R}}_{N_k}$ comprise the noise covariances.

This approach allows incorporating additional physical information on the process an its disturbances, in the form of constraints on the state and noise estimates:
\begin{alignat}{3}
    \mathbf{\hat{X}_{min} - C_{\hat{x}_{min}}} \varepsilon &\le \mathbf{\hat{X}} &&\le \mathbf{\hat{X}_{max} + C_{\hat{x}_{max}}} \varepsilon \\
    \mathbf{\hat{W}_{min} - C_{\hat{w}_{min}}} \varepsilon &\le \mathbf{\hat{W}} &&\le \mathbf{\hat{W}_{max} + C_{\hat{w}_{max}}} \varepsilon \\
    \mathbf{\hat{V}_{min} - C_{\hat{v}_{min}}} \varepsilon &\le \mathbf{\hat{V}} &&\le \mathbf{\hat{V}_{max} + C_{\hat{v}_{max}}} \varepsilon
\end{alignat}
and also $\varepsilon \ge 0$. The $\mathbf{\hat{X}}$ vector gathers the state estimates over the window. The $\mathbf{C}$ vectors are non-negative values that specify the softness of the associated bound, and $C$ globally weights the slack $\varepsilon$ (equal concern for relaxation). The problem \eqref{eq:J_MHE} is treated as a quadratic program for \texttt{LinModel}, and a nonlinear optimization, for \texttt{NonLinModel}. The optimal estimation problem is hard-coded as a single shooting formulation for now, but other transcription methods are planned for future releases.

\paragraph{\textnormal{\texttt{InternalModel}}}
Allows predictive control designs based on an internal model structure. It is based on the general approach of \citet{globPC}. The stochastic model of the unmeasured disturbances defaults to integrating white noise for each measured output (customizable). This is the equivalent of assuming that the disturbances are constant over the prediction horizon, similarly to dynamic matrix control (DMC). It supports asymptotically stable \texttt{LinModel} or \texttt{NonLinModel}.

\paragraph{\textnormal{\texttt{ManualEstimator}}}
Special subtype to disable the built-in state estimator and provide a custom feedback strategy.

\subsection{Predictive Controllers}

The prediction methodology applied throughout the package is mainly based on \citet{mpcMac} textbook. The three \texttt{PredictiveController} types are presented in the next sections.

\paragraph{\textnormal{\texttt{LinMPC}}}
Linear model predictive controller with soft or hard bound constraints. It minimizes the following objective function at each discrete time $k$:
\begin{multline}\label{eq:J_MPC}\!\!
J_{\mathit{MPC}} = 
    \mathbf{\big(\hat{R}_y - \hat{Y}\big)}^\intercal \mathbf{M}_{H_p} \mathbf{\big(\hat{R}_y - \hat{Y}\big)}   
    + \mathbf{\big(ΔU\big)}^\intercal \mathbf{N}_{H_c} \mathbf{\big(ΔU\big)} \\
    + \mathbf{\big(\hat{R}_u - U\big)}^\intercal \mathbf{L}_{H_p} \mathbf{\big(\hat{R}_u - U\big)} 
    + C \epsilon^2
\end{multline}
with the decision variables $\mathbf{ΔU}$ and $\epsilon$, the inputs increments over the control horizon $H_c$ and the slack variable, respectively. The vectors $\mathbf{\hat{Y}}$ and $\mathbf{\hat{R}_y}$ encompass the predictions of the outputs and their setpoints over the horizon $H_p$, respectively. The variables $\mathbf{U}$ and $\mathbf{\hat{R}_u}$ are similar but for the input setpoints. The matrices $\mathbf{M}_{H_p}$, $\mathbf{N}_{H_c}$ and $\mathbf{L}_{H_p}$ are Hermitian weights. In particular, if $\mathbf{M}_{H_p}$ is a block diagonal matrix, the last block allows specifying a terminal penalty different from the running weights.

The problem is subject to the following constraints:
\begin{alignat}{3}
    \mathbf{U_{min}  - C_{u_{min}}}  \epsilon 
         &\le \mathbf{U}  
        &&\le \mathbf{U_{max}  + C_{u_{max}}}  \epsilon \label{eq:u_constraints} \\
    \mathbf{ΔU_{min} - C_{Δu_{min}}} \epsilon 
         &\le \mathbf{ΔU} 
        &&\le \mathbf{ΔU_{max} + C_{Δu_{max}}} \epsilon \\
    \mathbf{Y_{min}  - C_{y_{min}}}  \epsilon 
         &\le \mathbf{\hat{Y}} 
        &&\le \mathbf{Y_{max}  + C_{y_{max}}}  \epsilon \\
    \mathbf{W_{min}  - C_{w_{min}}}  \epsilon 
         &\le \mathbf{W} 
        &&\le \mathbf{W_{max}  + C_{w_{max}}}  \epsilon \label{eq:w_constraints}
\end{alignat}
and also $\epsilon \ge 0$. The last line is for the custom linear inequality constraints. The vector $\mathbf{W}$ is defined as a linear combination of the inputs, outputs and setpoints. Box constraints on the terminal states are also possible: 
\begin{equation}
\mathbf{\hat{x}_{min}} - \mathbf{c_{\hat{x}_{min}}}\epsilon \le \mathbf{\hat{x}}_{i}(k+H_p) \le \mathbf{\hat{x}_{max}} + \mathbf{c_{\hat{x}_{max}}}\epsilon
\end{equation}
in which $i=k$ when using current estimators, and $i=k-1$ for delayed formulations. Note that time-varying constraints over the horizons and changing the bounds at runtime are supported (but not in the C codegen features).

The default optimizer is \texttt{OSQP.jl} that efficiently handles sparse problems \citep{osqp}, but the interface based on \texttt{JuMP.jl} allows switching among many quadratic solvers. For example, the dual active-set method of \citet{daqp} is sometimes more efficient on small and dense matrices. Testing this solver only takes 2 lines of code (see \cref{sec:successive_linearization} for an example). 

The user has the choice between a single shooting or a multiple shooting approach to transcribe the optimal control problem, which is also known as a condensed or sparse formulation, respectively. Move blocking with custom patterns for the manipulated inputs is also available to reduce the number of decision variables on long horizons. This object can be converted to a controller from \texttt{LinearMPC.jl} package for C code generation, but note that the solver and transcription is limited to the approach of \citet{daqp}.

\paragraph{\textnormal{\texttt{ExplicitMPC}}}
Linear model predictive controller without constraints, see \texttt{LinMPC} for the cost function. The computational costs are extremely low (the analytical solution of the quadratic problem leads to a single array division), therefore suitable for applications that require small sample times. It produces a control law similar to a finite-horizon linear-quadratic regulator (LQR), but with the control horizon $H_c$, the move suppression $\mathbf{N}_{H_c}$ and the input setpoint tracking $\mathbf{L}_{H_p}$ weights available as additional tuning parameters.

\paragraph{\textnormal{\texttt{NonLinMPC}}}
Nonlinear model predictive controller under constraints. The objective function includes an additional term for economic MPC:
\begin{equation}\label{eq:J_NMPC}
J_{\mathit{NMPC}} = J_{\mathit{MPC}} 
    + E J_E\big(\mathbf{U_e}, \mathbf{\hat{Y}_e}, \mathbf{\hat{D}_e}, \mathbf{p}, \epsilon \big)
\end{equation}
subject to the bound constraints \eqref{eq:u_constraints}--\eqref{eq:w_constraints} and the custom nonlinear inequality constraints:
\begin{equation}
\mathbf{g_c}\big(\mathbf{U_e}, \mathbf{\hat{Y}_e}, \mathbf{\hat{D}_e}, \mathbf{p}, \epsilon \big) \le \mathbf{0}
\end{equation}
with $J_{\mathit{MPC}}$ from \eqref{eq:J_MPC}. The user provides custom functions $J_E$ and $\mathbf{g_c}$ that returns the economic costs and the constraint values for a given set of inputs $\mathbf{U_e}$, outputs $\mathbf{\hat{Y}_e}$ and measured disturbances $\mathbf{\hat{D}_e}$, extended from $k$ to $k+H_p$ inclusively, as follows:
\begin{equation}
\mathbf{U_e} = 
\begin{bmatrix}
    \mathbf{U} \\ \mathbf{u}(k+H_p-1)
\end{bmatrix},\,\,
\mathbf{\hat{Y}_e} = 
\begin{bmatrix}
    \mathbf{\hat{y}}(k) \\ \mathbf{\hat{Y}}
\end{bmatrix},\,\,
\mathbf{\hat{D}_e} = 
\begin{bmatrix}
    \mathbf{d}(k) \\ \mathbf{\hat{D}}
\end{bmatrix} 
\end{equation} 
The parameter $\mathbf{p}$ includes arbitrary constants for computing the economical costs. 

The default optimizer is \texttt{Ipopt.jl}, an open-source interior point method developed by \citet{ipopt}. The user has the choice between shooting methods, trapezoidal collocation, likewise MATLAB, and orthogonal collocation using either Gauss-Radau or Gauss-Legendre quadrature, similarly to \citet{doMPC, gekko, apmonitor, acados}. The latter is particuliarly efficient on highly stiff dynamics. Except for single shooting, all these approaches are parallel-in-time, thus they all support optional multi-threading, which can be advantageous when the state-space functions are expensive to evaluate.
% !TeX encoding = UTF-8
% !TeX spellcheck = en_US
\section{Case Studies}
\label{sec:case_studies}

This section presents two case studies in simulation to demonstrate the design of linear and nonlinear MPCs. The objective here is to demonstrate how the package syntax facilitates the implementation and testing of predictive controllers and state estimators. The package manager of Julia eases the installation:
\begin{minted}{julia}
using Pkg; Pkg.add("ModelPredictiveControl")
\end{minted}
The first example details how to apply an MPC on a plant at each discrete time step. The second uses higher-level functionalities to quickly simulate closed-loop systems.

\subsection{Linear Design: Continuously Stirred Tank Reactor}

\begin{figure}[ht]
    \centering
    \begin{tikzpicture}[scale=1]
       
\fill[blue!10](0.0,0) rectangle (1.5,1.5);
\fill[blue!10](1.5,0) rectangle (1.8,0.3);

\draw[blue!30] (0.0, 1.5) -- (1.5,1.5);

\draw[<-,semithick] (0.45,0.75) arc (180:-150:0.3 and 0.3);

\draw[line width=2.2] (0, 2) -- (0, 0) -- (1.8,0);
\draw[line width=2.2] (1.8,0.3) -- (1.5,0.3) -- (1.5,2);

\pic[scale=0.6] (T) at (-0.6,1.20) {instrument=$y_T$};
\pic[scale=0.6] (L) at (+2.1,1.50) {instrument=$y_L$};

\draw [latex-latex] (1.675,1.25) -- (1.676,1.75);

\draw (T-right) -- ++(+0.6,0);
\draw (L-left)  -- ++(-0.3,0);

\pic [blue] (VC)    at (-0.6, 2.25) {valve=secondary};
\pic [blue] (VCact) at (-0.6, 2.25) {actuator};
\pic [red]  (VH)    at (2.1, 2.25) {valve=secondary};
\pic [red]  (VHact) at (2.1, 2.25) {actuator};

\draw[main stream, blue] (VC-right) -- node[above, black]{$u_c$} (0.5, 2.25) -| ++(0,-0.5);
\draw[main stream, red]  (VH-left)  -- node[above, black]{$u_h$} (1.0, 2.25) -| ++(0,-0.5);
\draw[main stream, blue] (-1.3, 2.25) -- (VC-left);
\draw[main stream, red]  (+2.8, 2.25) -- (VH-right);
\draw[main stream] (1.8,0.15) -- ++(0.5, 0);

\end{tikzpicture}
    \caption{CSTR process.}\label{fig:cstr}
\end{figure}
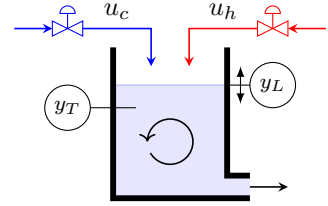

The example considers a continuously stirred tank reactor (CSTR) with a cold and hot water intakes, fed at a respective flow rate of $u_c$ and $u_h$. The manipulated input vector is thus $\mathbf{u} = [\begin{smallmatrix}u_c & u_h\end{smallmatrix}]^\intercal$. The liquid level $y_L$ and temperature $y_T$ constitute the measured output vector $\mathbf{y} = [\begin{smallmatrix}y_L & y_T\end{smallmatrix}]^\intercal$. \cref{fig:cstr} presents schematically the case study process.

\subsubsection{Linear Model}

At the steady-state operating point $u_c=u_h=20$, $y_L=50$, and $y_T=30$, the following linear model accurately describes the plant dynamics:
\begin{equation}
\mathbf{G}(s) = \frac{\mathbf{y}(s)}{\mathbf{u}(s)} =
\begin{bmatrix}
    \frac{1.90}{18s+1} & \frac{1.90}{18s+1} \\[3pt]
    \frac{-0.74}{8s+1} & \frac{0.74}{8s+1}
\end{bmatrix}
\end{equation}
The syntax to construct a linear model with a sample time of $\SI{2}{\second}$ is:
\begin{minted}{julia}
using ModelPredictiveControl, ControlSystemsBase
G = [ tf(1.90, [18, 1]) tf(1.90, [18, 1]);
      tf(-0.74,[8, 1])  tf(0.74, [8, 1]) ]
uop, yop = [20, 20], [50, 30]
vu , vy  = ["\$u_c\$", "\$u_h\$"], ["\$y_L\$", "\$y_T\$"]
model = setop!(LinModel(G, 2.0); uop, yop)
model = setname!(model; u=vu, y=vy)
\end{minted}
\spacerepl
\begin{minted}{julia-repl}
LinModel with a sample time Ts = 2.0 s:
└ dimensions:
  ├ 2 manipulated inputs u
  ├ 2 states x
  ├ 2 outputs y
  └ 0 measured disturbances d
\end{minted}
The \texttt{model} object is used for two purposes : to construct controllers and as a plant simulator to test the design. The \texttt{setname!} function modifies the $y$-axis labels in the plots.

\subsubsection{Linear Model Predictive Controller}

The objective is to maintain both the water temperature and level at their respective setpoints while constraining the level above 45:
\begin{minted}{julia}
mpc = setconstraint!(LinMPC(model), ymin=[45, -Inf])
\end{minted}
\spacerepl
\begin{minted}{julia-repl}
LinMPC controller with a sample time Ts = 2.0 s:
├ estimator: SteadyKalmanFilter
├ model: LinModel
├ optimizer: OSQP 
├ transcription: SingleShooting
└ dimensions:
  ├ 10 prediction steps Hp
  ├  2 control steps Hc
  ├  1 slack variable ϵ (control constraints)
  ├  2 manipulated inputs u (0 integrating states)
  ├  4 estimated states x̂
  ├  2 measured outputs ym (2 integrating states)
  ├  0 unmeasured outputs yu
  └  0 measured disturbances d
\end{minted}
By default, \texttt{LinMPC} controllers use \texttt{OSQP.jl} \citep{osqp} to solve the problem, soft constraints on output predictions $\mathbf{\hat y}$ for feasibility, and a \texttt{SteadyKalmanFilter} to estimate the plant states. The default predictive and control horizons are $H_p = 10 + n_k$ and $H_c = 2$, respectively, where $n_k$ is the number of delays in the linear model. The $H_c$ value is small by default to reduce the computational burden and the aggressiveness \citep{mpcHcAnalysis}. An attentive reader will also notice that the Kalman filter estimates two additional states compared to the plant model. These are the integrating states for the unmeasured plant disturbances, and they are automatically added at the model measured output $\mathbf{y^m}$ by default if observability is preserved.

Before closing the loop, the actual plant input $\mathbf{u}$ and measurement $\mathbf{y^m}$ should initialize the estimates $\mathbf{\hat{x}}$ at the steady-state solution that leads to $\mathbf{\hat{y}^m}(0)=\mathbf{y^m}(0)$. This approach results in a bumpless transfer, especially with estimators in the predictor form. The \texttt{initstate!} function finds this solution for \texttt{LinModel}. Since the plant is the model here, its output initializes the states. \texttt{LinModel} objects are callable for this purpose. Once done, imposing step changes on the setpoint \texttt{ry} and on a load disturbance \texttt{ul} tests the closed-loop performance of \texttt{mpc}:
\begin{minted}{julia}
function test_mpc(mpc, plant)
    plant.x0 .= 0; y = plant() # or evaloutput(plant)
    initstate!(mpc, plant.uop, y)
    N = 75; ry = [50, 30]; ul = 0
    U, Y, Ry = zeros(2, N), zeros(2, N), zeros(2, N)
    for i = 1:N
        i == 26 && (ry = [48, 35])
        i == 51 && (ul = -10)
        y = plant() # simulated measurements
        preparestate!(mpc, y) # prepare mpc estimate
        u = mpc(ry) # or moveinput!(mpc, ry)
        U[:,i], Y[:,i], Ry[:,i] = u, y, ry
        updatestate!(mpc, u, y) # update mpc estimate
        updatestate!(plant, u+[0,ul]) # update simulator
    end
    return U, Y, Ry
end
U_data, Y_data, Ry_data = test_mpc(mpc, model)
\end{minted}
Calling \texttt{LinMPC} instances computes and returns $\mathbf{u}(k)$. It is worth mentioning that additional information like the optimal output predictions $\mathbf{\hat{Y}}$ can be retrieved by calling \texttt{getinfo} after solving the problem. Also, calling the \texttt{preparestate!} method on \texttt{mpc} prepares the state estimates for the current control period, and \texttt{updatestate!} updates them for the next one (the same logic applies for \texttt{plant}). Lastly, the package implements a \texttt{Plots.jl} recipe \citep{plots} to quickly visualize the results:
\begin{minted}{julia}
res = SimResult(mpc, U_data, Y_data; Ry_data)
using Plots; plot(res)
\end{minted}

\begin{figure}[h]
    \centering
    \includegraphics[width=\columnwidth]{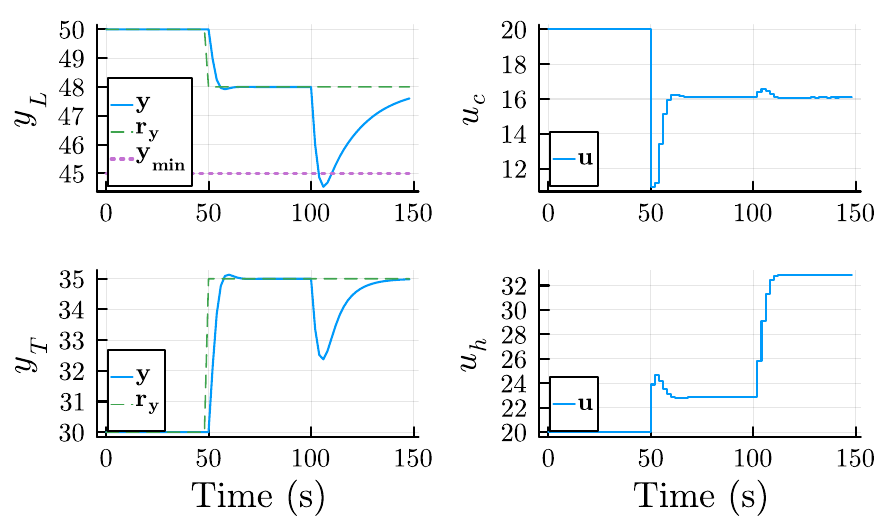}
    \caption{CSTR closed-loop simulation (MPC).}\label{fig:plot_LinMPC1}
\end{figure}

\cref{fig:plot_LinMPC1} shows that the controller violates the constraints around 110 s because of the disturbance. Adding feedforward compensation can mitigate this.

\subsubsection{Feedforward Compensation}

Suppose that the load disturbance $u_l$ is in fact caused by a separate hot water pipe that discharges into the tank. Measuring this flow rate allows us to incorporate feedforward compensation with $\mathbf{d}=u_l$. The new model is:

\begin{minted}{julia}
model_d = LinModel([G G[1:2, 2]], 2.0; i_d=[3])
model_d = setop!(model_d; uop, yop, dop=[20])
model_d = setname!(model_d; u=vu, y=vy, d=["\$u_l\$"])
\end{minted}
\spacerepl
\begin{minted}{julia-repl}
LinModel with a sample time Ts = 2.0 s:
└ dimensions:
  ├ 2 manipulated inputs u
  ├ 4 states x
  ├ 2 outputs y
  └ 1 measured disturbances d
\end{minted}
The simulation requires a new \texttt{LinMPC} instance based on \texttt{model\_d} and a new test function that explicitly employs the current disturbance measurement:
\begin{minted}{julia}
mpc_d = setconstraint!(LinMPC(model_d), ymin=[45, -Inf])
function test_mpc_d(mpc_d, plant)
    plant.x0 .= 0; y = plant(); d = [20]
    initstate!(mpc_d, plant.uop, y, d)
    N = 75; ry = [50, 30]; ul = 0
    U, Y, Ry = zeros(2, N), zeros(2, N), zeros(2, N)
    for i = 1:N
        i == 26 && (ry = [48, 35])
        i == 51 && (ul = -10)
        y, d = plant(), [20+ul] # simulated measurements
        preparestate!(mpc_d, y, d) # d in arguments
        u = mpc_d(ry, d) # d in arguments
        U[:,i], Y[:,i], Ry[:,i] = u, y, ry
        updatestate!(mpc_d, u, y, d) # d in arguments
        updatestate!(plant, u+[0,ul])
    end
    return U, Y, Ry
end
U_data, Y_data, Ry_data = test_mpc_d(mpc_d, model)
res = SimResult(mpc, U_data, Y_data; Ry_data)
plot(res)
\end{minted}

\begin{figure}[h]
    \centering
    \includegraphics[width=\columnwidth]{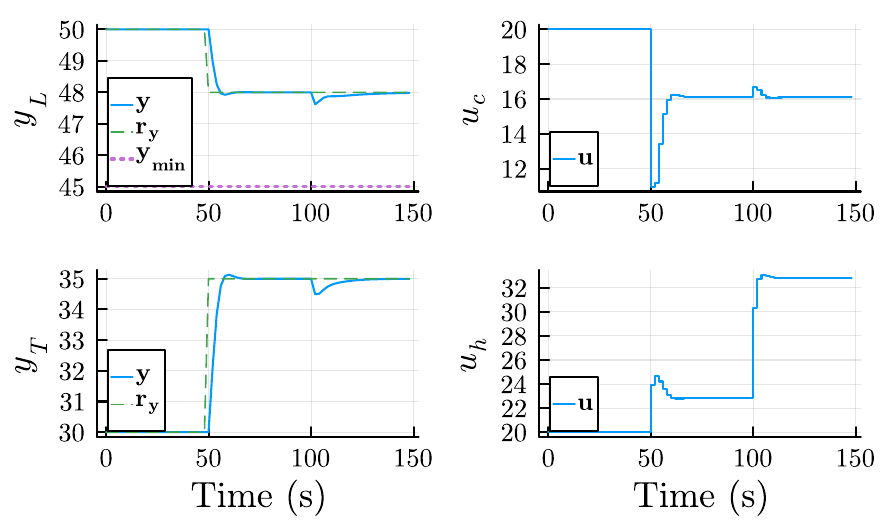}
    \caption{CSTR closed-loop simulation with feedforward (MPC).}\label{fig:plot_LinMPC2}
\end{figure}

\cref{fig:plot_LinMPC2} shows that the feedforward compensation handles the disturbance without violating the constraint. Note that measured disturbances are assumed constant in the future by default but custom $\mathbf{\hat{D}}$ predictions are possible. The same applies for the setpoints $\mathbf{\hat{R}_y}$ and $\mathbf{\hat{R}_u}$. For instance, the stochastic predictions of an \texttt{InternalModel} object could generate these two vectors.

\subsection{Nonlinear Design : Inverted Pendulum}
\label{sec.nonlinear_design}

\begin{figure}[ht]
    \centering
    \begin{tikzpicture}[scale=1]
       
\fill (-0.85,-0.2) rectangle (0.85,0.2);
\draw (-1.5,0) -- (1.5,0) node[above, very near start]{$\tau$};
\draw[fill=white] (0,0) circle (0.75);

\fill (0,0) circle (0.15);

\draw[line width=2.2] (0,0) -- (0,1.3);
\draw[densely dotted, gray] (0,0) -- (0,-0.5);
\draw[gray, <-] (0,0.3) arc [start angle=90, end angle=-90, radius=0.3];
\node[gray] at (0.5, 0.0) {$\theta$};

\node[black] at (-0.35, 0.0) {$K$};

\filldraw (0, 1.6) circle (0.3) node[white]{$m$};

\draw [Stealth-Stealth] (1.2, 0) -- node[fill=white]{$L$} (1.2, 1.6);
\draw (1.15, 1.6) -- (1.25, 1.6);

\end{tikzpicture}
    \caption{Inverted pendulum.}\label{fig:pendulum}
\end{figure}
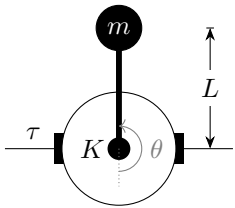

For this case study, the goal is to control the angular position $\theta$ of a pendulum attached to a motor, thus $\mathbf{y} = \theta$. The manipulated input is the motor torque $\mathbf{u} = \tau$ in \si{\newton\meter}. \cref{fig:pendulum} depicts the system. 

\subsubsection{Nonlinear Model}

The model is:
\begin{align}
\dot{\theta}(t) &= \omega(t) \\
\dot{\omega}(t) &= -\frac{g}{L}\sin\big(\theta(t)\big) -\frac{K}{m}\omega(t) + \frac{1}{m L^2}\tau(t) \label{eq.pendulum_speed}
\end{align}
in which $g$ is the gravitational acceleration in \si{\meter\per\second\squared}, $L$, the pendulum length in \si{\meter}, $K$, the friction coefficient at the pivot point in \si{\kilogram\per\second}, and $m$, the mass attached at the end of the pendulum in \si{\kilogram}. They are all bundled in the parameter vector $\mathbf{p}=[\begin{smallmatrix}g & L & K & m\end{smallmatrix}]'$. By default, the \texttt{NonLinModel} constructor assumes continuous-time equations and discretizes them with Runge-Kutta:
\begin{minted}{julia}
using ModelPredictiveControl
function f!(ẋ, x, u, _ , p)
    g, L, K, m = p       # [m/s²], [m], [kg/s], [kg]
    θ, ω = x[1], x[2]    # [rad], [rad/s]
    τ = u[1]             # [Nm]
    ẋ[1] = ω
    ẋ[2] = -g/L*sin(θ) - K/m*ω + τ/m/L^2
end
h!(y, x, _ , _ ) = (y[1] = 180/π*x[1])   # [°]
p = [9.8, 0.4, 1.2, 0.3]
nu = 1; nx = 2; ny = 1; Ts = 0.1
model = NonLinModel(f!, h!, Ts, nu, nx, ny)
vu = ["\$τ\$ (Nm)"]
vx = ["\$θ\$ (rad)", "\$ω\$ (rad/s)"]
vy = ["\$θ\$ (°)"]
model = setname!(model; u=vu, x=vx, y=vy)
\end{minted}
\spacerepl\vspace{0.6em}
\begin{minted}{julia-repl}
NonLinModel with a sample time Ts = 0.1 s:
├ solver: RungeKutta(4)
├ jacobian: AutoForwardDiff
└ dimensions:
  ├ 1 manipulated inputs u
  ├ 2 states x
  ├ 1 outputs y
  └ 0 measured disturbances d
\end{minted}
The output function $\mathbf{h}$ converts the $\theta$ angle to degrees. The plant model relies on mutating state-space functions to reduce the memory allocations and the computational burden as well. The more intuitive non-mutating syntax can still be used by providing functions with one less argument, e.g.: \texttt{f(x,u,d,p)=x+u+d+p} and \texttt{h(x,d,p)=x+d+p}.

\subsubsection{Nonlinear State Estimator}

The state estimates of an \texttt{UnscentedKalmanFilter} will feed the controller:
\begin{minted}{julia}
σQ = [0.1, 1.0]; σR=[5.0]; nint_u=[1]; σQint_u=[0.1]
estim = UnscentedKalmanFilter(model; σQ, σR, 
                              nint_u, σQint_u)
\end{minted}
\spacerepl
\begin{minted}{julia-repl}
UnscentedKalmanFilter estimator with a sample time Ts =
0.1 s:
├ model: NonLinModel
└ dimensions:
  ├ 1 manipulated inputs u (1 integrating states)
  ├ 3 estimated states x̂
  ├ 1 measured outputs ym (0 integrating states)
  ├ 0 unmeasured outputs yu
  └ 0 measured disturbances d
\end{minted}
The vectors \texttt{σQ} and \texttt{σR} are the standard deviations of the process and sensor noises, respectively. The value for the velocity $\omega$ is higher here (\texttt{σQ} second value) since \eqref{eq.pendulum_speed} includes an uncertain parameter: the friction coefficient $K$. Also, the argument \texttt{nint\_u} explicitly adds one integrating state at the model input, the motor torque $\tau$, with an associated standard deviation \texttt{σQint\_u} of $\SI{0.1}{\newton\meter}$. On this subject, custom models for the unmeasured disturbances is also possible by constructing a \texttt{NonLinModel} with the state-space functions based on the augmented dynamics directly, and by specifying \texttt{nint\_u=0} and \texttt{nint\_ym=0} when constructing the estimator. The estimator tuning is tested on a plant with a \SI{25}{\percent} larger friction coefficient $K$: 
\begin{minted}{julia}
p_plant = copy(p); p_plant[3] = 1.25*p[3]
plant = NonLinModel(f!, h!, Ts, nu, nx, ny; p=p_plant)
N = 35; u = [0.5]; 
res = sim!(estim, N, u; plant, y_noise=[0.5])
using Plots; plot(res, plotu=false, plotxwithx̂=true)
\end{minted}

\begin{figure}[h]
    \centering
    \includegraphics[width=\columnwidth]{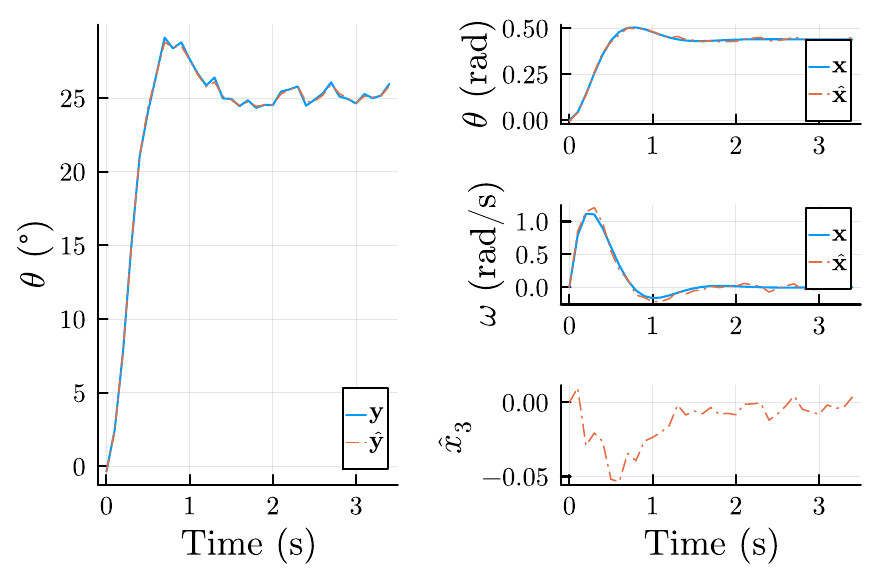}
    \caption{Pendulum state estimation.}\label{fig:plot_NonLinMPC1}
\end{figure}

\cref{fig:plot_NonLinMPC1} indicates that the Kalman filter performance seems sufficient for control. The estimate $\hat{x}_3$ is the integrating state on the torque $\tau$ that compensates for static errors. 

\subsubsection{Nonlinear Model Predictive Controller (NMPC)}

As the motor torque is limited to \num{-1.5} to \SI{1.5}{\newton\meter}, the input constraints are incorporated to a \texttt{NonLinMPC} object:
\begin{minted}{julia}
Hp, Hc, Mwt, Nwt, Cwt = 20, 2, [0.5], [2.5], Inf
transcription = MultipleShooting()
nmpc = NonLinMPC(estim; Hp, Hc, Mwt, 
                 Nwt, Cwt, transcription)
umin, umax = [-1.5], [+1.5]
nmpc = setconstraint!(nmpc; umin, umax)
\end{minted}
\spacerepl\vspace{0.4em}
\begin{minted}{julia-repl}
NonLinMPC controller with a sample time Ts = 0.1 s:
├ estimator: UnscentedKalmanFilter
├ model: NonLinModel
├ optimizer: Ipopt 
├ transcription: MultipleShooting
├ gradient: AutoForwardDiff
├ jacobian: AutoSparse (AutoForwardDiff, TracerSparsity...
├ hessian: nothing
└ dimensions:
  ├ 20 prediction steps Hp
  ├  2 control steps Hc
  ├  0 slack variable ϵ (control constraints)
  ├  1 manipulated inputs u (1 integrating states)
  ├  3 estimated states x̂
  ├  1 measured outputs ym (0 integrating states)
  ├  0 unmeasured outputs yu
  └  0 measured disturbances d
\end{minted}
The keyword arguments \texttt{Mwt} and \texttt{Nwt} are the output setpoint tracking and move suppression weights, respectively. The option \texttt{Cwt=Inf} disables the slack variable $\epsilon$ when not needed. Here, a \texttt{MultipleShooting} method transcribes the optimization problem to ease convergence on the highly nonlinear and unstable pendulum model, but also to reproduce MATLAB's internals for nonlinear MPC, which is hard-coded as a multiple shooting. 

By default, the package \texttt{ForwardDiff.jl} computes the gradient and the constraint Jacobians, and the Hessian of the Lagrangian is skipped, thus falling back to the internal quasi-Newton approximation of \texttt{Ipopt.jl}. With non-single-shooting transcriptions, the sparsity of the Jacobian (and Hessian, if enabled) is automatically detected and exploited to reduce the computational burden, thanks to the \texttt{DifferentiationInterface.jl} dependency \citep{differentiationinterface}. Several configurations related to the sparsity pattern are also available, as well as the option to use finite differences, similarly to MATLAB.  

An angular setpoint of \SI{180}{\degree} (inverted position) tests \texttt{mpc} performances on the plant:
\begin{minted}{julia}
x_0 = [0, 0]; x̂_0 = [0, 0, 0]; ry = [180]
res_r = sim!(nmpc, N, ry; plant, x_0, x̂_0)
plot(res_r)
\end{minted}

\begin{figure}[ht]
    \centering
    \includegraphics[width=\columnwidth]{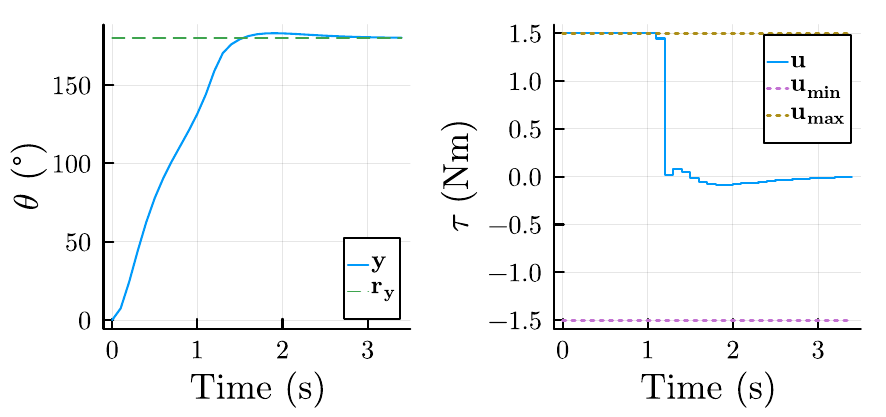}
    \caption{Pendulum output setpoint tracking (NMPC).}\label{fig:plot_NonLinMPC2}
\end{figure}

The result in \cref{fig:plot_NonLinMPC2} suggests that the controller is robust enough to variations on the $K$ coefficient. 

Starting from this inverted position, \cref{fig:plot_NonLinMPC3} shows the closed-loop response to a step disturbances of \SI{10}{\degree} is satisfactory:
\begin{minted}{julia}
x_0 = [π, 0]; x̂_0 = [π, 0, 0]; y_step = [10]
res_d = sim!(nmpc, N, [180.0]; plant, x_0, x̂_0, y_step)
plot(res_d)
\end{minted}

\begin{figure}[ht]
    \centering
    \includegraphics[width=\columnwidth]{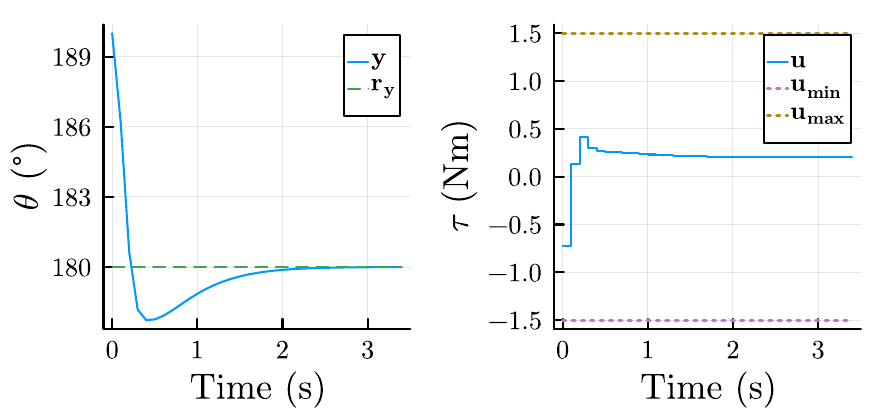}
    \caption{Pendulum output disturbance rejection (NMPC).}\label{fig:plot_NonLinMPC3}
\end{figure}

The next section presents an economic model predictive controller on the pendulum.

\subsubsection{Economic model predictive control (EMPC)}

For this case study, the controller will aim to reduce the energy consumed by the motor. The power in watts transmitted by the motor to the pendulum is:
\begin{equation}
P(t) = \tau(t) \omega(t) 
\end{equation}
Thus, the work in joule done by the motor from $t_0$ to $t_{end}$ is:
\begin{equation}
W = \int_{t_0}^{t_{end}} P(t) \mathrm{d}t 
\end{equation}
With the sampling time $T_s$ in seconds and the prediction horizon $H_p$ in time steps, computing the integral with the left-endpoint rectangle method and defining the limits as $t_0=k T_s$ and $t_{end} = (k + H_p) T_s$ give:
\begin{equation}\label{eq:pendulum_W}
W \approx T_s \sum_{j = 0}^{H_p - 1} \tau(k+j) \omega(k+j) 
\end{equation}

The objective will now include an additive term that penalizes the work done by the motor $W$ to reduce the energy consumption. Notice that \eqref{eq:pendulum_W} is a function of the manipulated input $\tau$ and the angular speed $\omega$, a state that is not measured (only the angle $\theta$ is measured here). As the arguments of the economic function $J_E$ in \eqref{eq:J_NMPC} do not include the states, the speed is now defined as an unmeasured output to design a Kalman filter similar to the previous one ($\mathbf{y^m} = \theta$ and $\mathbf{y^u} = \omega$): 

\begin{minted}{julia}
h2!(y, x, _ , _ ) = (y[1] = 180/π*x[1]; y[2]=x[2])
nu, nx, ny = 1, 2, 2
model2 = NonLinModel(f!, h2!, Ts, nu, nx, ny; p)
plant2 = NonLinModel(f!, h2!, Ts, nu, nx, ny; p=p_plant)
model2 = setname!(model2, u=vu, x=vx, y=[vy; vx[2]])
plant2 = setname!(plant2, u=vu, x=vx, y=[vy; vx[2]])
estim2 = UnscentedKalmanFilter(model2; σQ, σR, 
                               nint_u, σQint_u, i_ym=[1])
\end{minted}
The \texttt{plant2} object based on \texttt{h2!} is also required since \texttt{sim!} expects that the plant output vector $\mathbf{y}$ corresponds to the controller model output vector. Now, these lines define the $J_E$ function and the controller:
\begin{minted}{julia}
function JE(UE, ŶE, _ , p , _ )
    Ts = p
    τ, ω = UE[1:end-1], ŶE[2:2:end-1]
    return Ts*sum(τ.*ω)
end
p = Ts; Mwt2 = [Mwt; 0.0]; Ewt = 3.5e3
empc = NonLinMPC(estim2; Hp, Hc, Mwt=Mwt2, 
                 Nwt, Cwt, JE, Ewt, p, transcription)
empc = setconstraint!(empc; umin, umax)
\end{minted}
\spacerepl
\begin{minted}{julia-repl}
NonLinMPC controller with a sample time Ts = 0.1 s:
├ estimator: UnscentedKalmanFilter
├ model: NonLinModel
├ optimizer: Ipopt 
├ transcription: MultipleShooting
├ gradient: AutoForwardDiff
├ jacobian: AutoSparse (AutoForwardDiff, TracerSparsity...
├ hessian: nothing
└ dimensions:
  ├ 20 prediction steps Hp
  ├  2 control steps Hc
  ├  0 slack variable ϵ (control constraints)
  ├  1 manipulated inputs u (1 integrating states)
  ├  3 estimated states x̂
  ├  1 measured outputs ym (0 integrating states)
  ├  1 unmeasured outputs yu
  └  0 measured disturbances d
\end{minted}

The keyword argument \texttt{Ewt} is the $E$ weight in \eqref{eq:J_NMPC}. The economic term must be large enough to be significant but a too high value can lead to a static error on the angle setpoint. The second element of \texttt{Mwt2} is zero since the speed $\omega$ is not requested to track a setpoint. \cref{fig:plot_EconomMPC1} shows that closed-loop response to a \SI{180}{\degree} setpoint is similar:
\begin{minted}{julia}
x_0 = [0, 0]; x̂_0 = [0, 0, 0]; ry = [180; 0]
res2_r = sim!(empc, N, ry; plant=plant2, x_0, x̂_0)
plot(res2_r, ploty=[1])
\end{minted}

\begin{figure}[h]
    \centering
    \includegraphics[width=\columnwidth]{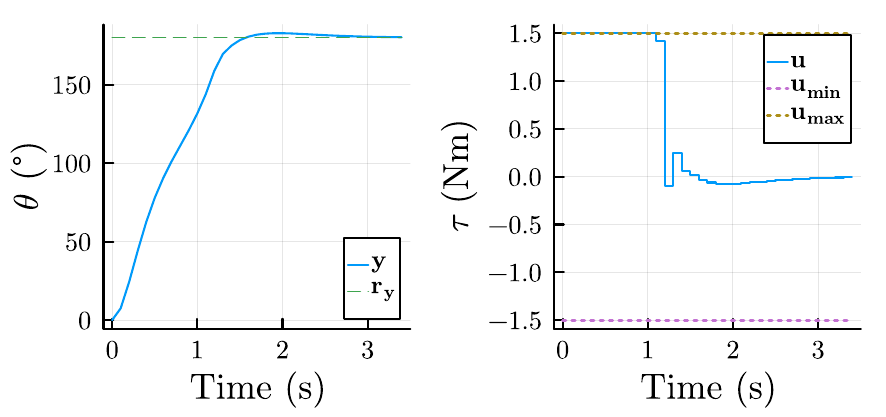}
    \caption{Pendulum output setpoint tracking (EMPC).}\label{fig:plot_EconomMPC1}
\end{figure}

\noindent and the energy consumption is slightly lower:
\begin{minted}{julia}
function calcW(res)
    τ, ω = res.U_data[1, 1:end-1], res.X_data[2, 1:end-1]
    return Ts*sum(τ.*ω)
end
Dict(:W_nmpc => calcW(res_r), :W_empc => calcW(res2_r))
\end{minted}
\spacerepl\vspace{0.3em}
\begin{minted}{julia-repl}
Dict{Symbol, Float64} with 2 entries:
  :W_empc => 3.90053
  :W_nmpc => 3.92078
\end{minted}

\noindent Also, for a \SI{10}{\degree} step disturbance:
\begin{minted}{julia}
x_0 = [π, 0]; x̂_0 = [π, 0, 0]; y_step = [10; 0]
res2_d = sim!(empc, N, ry; plant=plant2, x_0, x̂_0, y_step)
plot(res2_d, ploty=[1])
\end{minted}

\begin{figure}[h]
    \centering
    \includegraphics[width=\columnwidth]{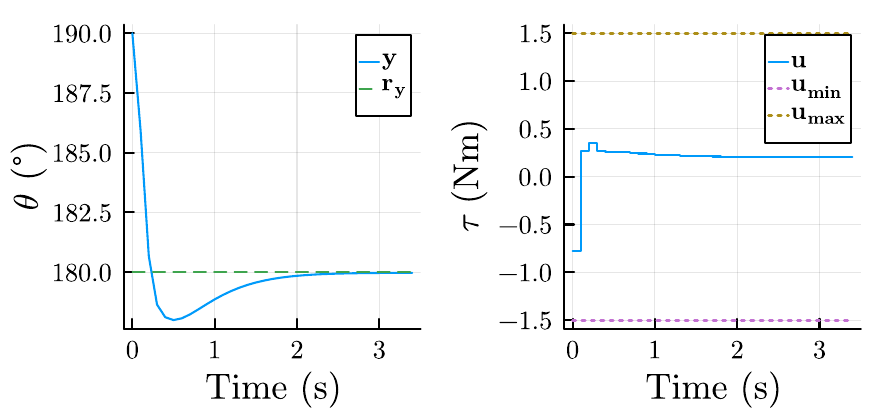}
    \caption{Pendulum output disturbance rejection (EMPC).}\label{fig:plot_EconomMPC2}
\end{figure}

\noindent \cref{fig:plot_EconomMPC2} shows that the new controller is able to recuperate a little more energy from the pendulum (i.e. negative work):
\begin{minted}{julia}
Dict(:W_nmpc => calcW(res_d), :W_empc => calcW(res2_d))
\end{minted}
\spacerepl\vspace{0.3em}
\begin{minted}{julia-repl}
Dict{Symbol, Float64} with 2 entries:
  :W_empc => -0.0542454
  :W_nmpc => -0.0440395
\end{minted}
Of course, this gain is only exploitable if the motor electronic includes some kind of regenerative circuitry.

\subsubsection{Successive Linearization Model Predictive Control (SLMPC)}
\label{sec:successive_linearization}

The \texttt{setmodel!} method allows online adaptation of a linear plant model. Combined with the automatic linearization of \texttt{linearize}, based on \texttt{ForwardDiff.jl} by default, a successive linearization MPC can be designed with minimal efforts. But first, the default \texttt{LinMPC} optimizer (\texttt{OSQP.jl}) does not perform as well as \texttt{DAQP.jl} on unstable plants like here \citep{daqp}. The syntax to install it is: 
\begin{minted}{julia}
using Pkg; Pkg.add(["JuMP","DAQP"])
using JuMP, DAQP
optim = JuMP.Model(DAQP.Optimizer, add_bridges=false)
\end{minted}
The bridges are a concept introduced by \texttt{JuMP.jl} to convert the constraints into a form that is natively supported by a specific solver (e.g.: from inequality to equality constraints). Disabling them slightly reduce the overhead when it is not required, like here. Next, the default asymptotic Kalman Filter does not support \texttt{setmodel!}, so we need to use the time-varying \texttt{KalmanFilter} instead:
\begin{minted}{julia}
linmodel = linearize(model, x=[0, 0], u=[0])
kf = KalmanFilter(linmodel; σQ, σR, nint_u, σQint_u)
mpc3 = LinMPC(kf; Hp, Hc, Mwt, Nwt, Cwt, optim)
mpc3 = setconstraint!(mpc3; umin, umax)
\end{minted}
\spacerepl
\begin{minted}{julia-repl}
LinMPC controller with a sample time Ts = 0.1 s:
├ estimator: KalmanFilter
├ model: LinModel
├ optimizer: DAQP 
├ transcription: SingleShooting
└ dimensions:
  ├ 20 prediction steps Hp
  ├  2 control steps Hc
  ├  0 slack variable ϵ (control constraints)
  ├  1 manipulated inputs u (1 integrating states)
  ├  3 estimated states x̂
  ├  1 measured outputs ym (0 integrating states)
  ├  0 unmeasured outputs yu
  └  0 measured disturbances d
\end{minted}
A custom simulation function that successively linearizes the pendulum model and updates the \texttt{LinMPC} controller with \texttt{setmodel!} is required: 
\begin{minted}{julia}
function sim2!(mpc, nlmodel, N, ry, plant, x, x̂, y_step)
    U, Y, Ry = zeros(1, N), zeros(1, N), zeros(1, N)
    setstate!(plant, x); setstate!(mpc, x̂)
    initstate!(mpc, [0], plant())
    linmodel = linearize(nlmodel; u=[0], x=x̂[1:2])
    setmodel!(mpc, linmodel)
    for i = 1:N
        y = plant() + y_step
        x̂ = preparestate!(mpc, y)
        u = mpc(ry)
        linearize!(linmodel, nlmodel; u, x=x̂[1:2])
        setmodel!(mpc, linmodel) 
        U[:,i], Y[:,i], Ry[:,i] = u, y, ry
        updatestate!(mpc, u, y)
        updatestate!(plant, u)
    end
    U_data, Y_data, Ry_data = U, Y, Ry
    return SimResult(mpc, U_data, Y_data; Ry_data)
end
\end{minted}
The new model is set after solving the optimization problem, and before updating the state estimate, that is, when both $\mathbf{u}(k)$ and $\mathbf{\hat{x}}_k(k)$ are available as the new operating point. As shown in \cref{fig:plot_SuccLinMPC1} and \ref{fig:plot_SuccLinMPC2}, the SLMPC closed-loop response is similar to the nonlinear MPC, both for the 180° setpoint:
\begin{minted}{julia}
x_0 = [0, 0]; x̂_0 = [0, 0, 0]; ry = [180]
res3_r = sim2!(mpc3, model, N, ry, plant, x_0, x̂_0, [0])
plot(res3_r)
\end{minted}

\begin{figure}[h]
    \centering
    \includegraphics[width=\columnwidth]{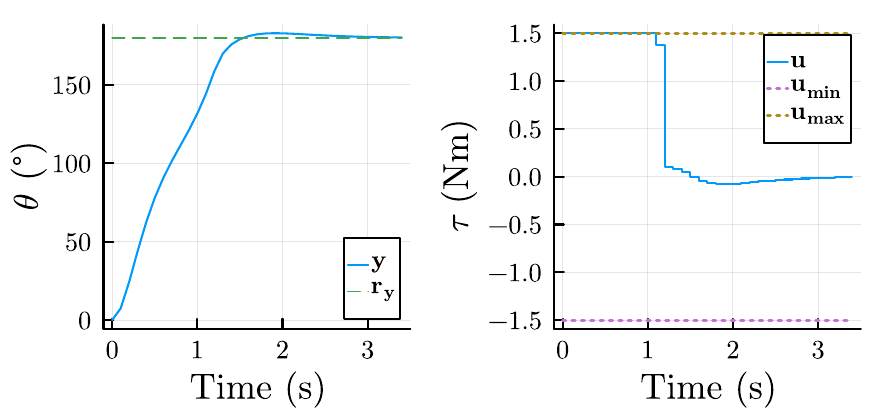}
    \caption{Pendulum output setpoint tracking (SLMPC).}\label{fig:plot_SuccLinMPC1}
\end{figure}

\noindent and the 10° step disturbance:
\begin{minted}{julia}
x_0 = [π, 0]; x̂_0 = [π, 0, 0]; ry = [180]
res3_d = sim2!(mpc3, model, N, ry, plant, x_0, x̂_0, [10])
plot(res3_d)
\end{minted}

\begin{figure}[h]
    \centering
    \includegraphics[width=\columnwidth]{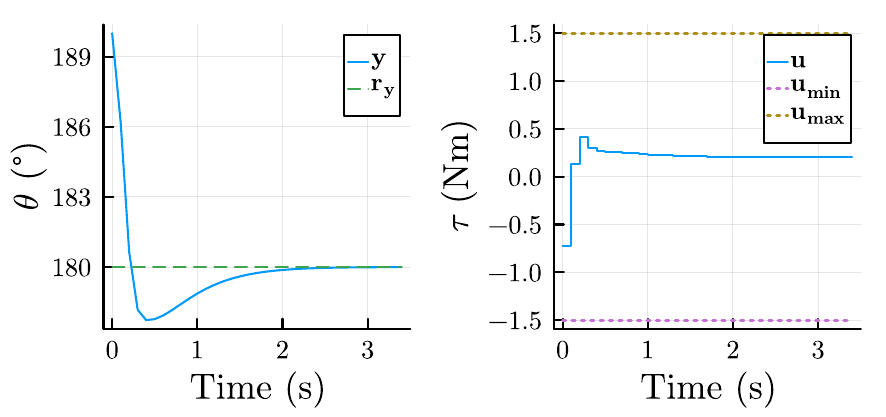}
    \caption{Pendulum output disturbance rejection (SLMPC).}\label{fig:plot_SuccLinMPC2}
\end{figure}

The next section presents the interesting result: the solving time is drastically lower.

\subsection{Benchmarks}
\label{sec:benchmarks}

The time required to simulate the previous case studies is a benchmark of MPC performance in terms of computational cost (the cost to simulate the plant being negligible). The case studies are reproduced on MATLAB to compare the performance of the two frameworks. A special care is taken to ensure that that the closed-loop solution in MATLAB is identical to Julia’s counterpart. A comparison with other MPC toolkits, for instance the various open source packages in Python and C/C++, is out of the scope of this paper.

The median simulation time of 500 simulations, for MPC and SLMPC, and 50 simulations, for NMPC and EMPC, provide a performance benchmark robust to outliers. It must be emphasized that the reported values are the total time to simulate the whole scenario, that is $N=75$ time steps for the CSTR, and $N=35$, for the pendulum.

The MATLAB implementations use the model predictive control toolbox. Additional licenses to the control system and optimization toolboxes are needed to design and execute MPCs. Note that the SLMPC code relies on the symbolic toolbox, since there is no automatic differentiation in MATLAB. All the scripts are executed as is, in the interpreter. Code generation is typically required to produce fast dedicated code. But, as mentioned in the introduction, it creates other problems like extra licensing fees (e.g.: MATLAB/Simulink Coder). Also note that MATLAB only supports the less efficient non-mutating syntax for the nonlinear state-space functions.

Since the performance is solver-dependent, multiple optimization algorithms are tested. The results include four types of quadratic/nonlinear solvers. The operator-splitting (OS) methods are: 
\begin{itemize}
    \item \texttt{OSQP} package in Julia \citep{osqp}
    \item \texttt{admm} configuration in MATLAB \citep{admm}
\end{itemize}
For active-set (AS) methods, the optimizers are: 
\begin{itemize}
    \item \texttt{DAQP} package in Julia \citep{daqp}
    \item \texttt{active-set} configuration in MATLAB \citep{activeset}
\end{itemize}
For interior-point (IP) methods, the solvers are: 
\begin{itemize}
    \item \texttt{Ipopt} package in Julia \citep{ipopt}
    \item \texttt{interior-point} configuration in MATLAB \citep{interiorpoint}
\end{itemize}
And, lastly, the sequential quadratic (SQ) methods are:
\begin{itemize}
    \item \texttt{UnoSolver} package in Julia with \texttt{filtersqp} and a line search globalization \citep{uno}
    \item \texttt{sqp} configuration in MATLAB \citep{sqp}
\end{itemize}

All the solvers in Julia are free and open-source. A separate keyword argument in the controller constructors allows customizing the solver (see \cref{sec:successive_linearization} for an example). Only the active-set results are include for the SLMPC since \texttt{OSQP.jl} does not work as well on the pendulum (without major modifications in the optimizer settings).

A laptop with an Intel\textsuperscript{\textregistered} Core\textsuperscript{\texttrademark} i7-1165G7 (8 cores at \SI{4.70}{\giga\hertz}) and a Linux v6.17.0 kernel runs the benchmarks. The scripts are executed on Julia v1.12.6, the v2.3.1 release of this package and MATLAB R2024a. A public repository provides the code to reproduce the results, see \url{https://github.com/franckgaga/mpcPackageJulia.tex}.
It can be cloned, and by executing the following commands in Julia with the git root folder as the working directory:
\begin{minted}{julia}
using Pkg; Pkg.activate("julia"); Pkg.instantiate()
\end{minted}
an identical environment with the same versions of the packages will be created.

\begin{table}[t]
    \centering
    \caption{Julia and MATLAB Benchmarks on the Case Studies.}
    \label{tab:benchamrks}
    \vspace{0.5em}
    \centering
    \footnotesize
    \begin{tabular}{llllrr}
	
\toprule %=======================================================================

	  &	& & & \multicolumn{2}{c}{Median Time (s)} \\ \cmidrule(l){5-6}
Plant & Control & Test & Solver & Julia & MATLAB \\
\midrule %--------------------------------------------------------------------

CSTR		& MPC	& W/o $\mathbf{d}$	& OS & \num{0.0012} & \num{0.0205}	\\
CSTR		& MPC	& W/o $\mathbf{d}$	& AS & \num{0.0032} & \num{0.0182}	\\
CSTR		& MPC	& With $\mathbf{d}$ & OS & \num{0.0012} & \num{0.0201}	\\
CSTR		& MPC	& With $\mathbf{d}$ & AS & \num{0.0034} & \num{0.0176}	\\
Pendulum 	& NMPC	& Track. 	   		& IP & \num{0.2621} & \num{1.3989}	\\
Pendulum 	& NMPC	& Track. 	   		& SQ & \num{0.1650} & \num{0.5924}	\\
Pendulum    & NMPC	& Regul. 			& IP & \num{0.2360} & \num{1.5525} 	\\
Pendulum    & NMPC	& Regul. 			& SQ & \num{0.1677} & \num{0.5691} 	\\
Pendulum    & EMPC	& Track.			& IP & \num{0.2994} & \num{1.1633} 	\\
Pendulum    & EMPC	& Track.			& SQ & \num{0.1805} & \num{0.6966} 	\\
Pendulum	& EMPC	& Regul. 			& IP & \num{0.3324} & \num{0.8827} 	\\
Pendulum	& EMPC	& Regul. 			& SQ & \num{0.1710} & \num{0.6279}  \\
Pendulum    & SLMPC & Track.			& AS & \num{0.0072} & \num{0.0264}  \\
Pendulum    & SLMPC & Regul.			& AS & \num{0.0072} & \num{0.0263}  \\
	
\bottomrule %====================================================================
	
\end{tabular}
\end{table}

The benchmarks are presented in \cref{tab:benchamrks}. The linear MPC is significantly faster in Julia, approximately from 5 to 17 times faster. This result is however amplified by \texttt{OSQP.jl}, a modern and fast OS quadratic optimizer, which seems to perform considerably better than MATLAB's counterpart (\texttt{admm}). But the authors would argue that it is a consequence of \texttt{ModelPredictiveControl.jl} modular design. Any modern optimization tools can be leveraged with minimal effort and without any package update, presuming that they offer a \texttt{JuMP.jl} interface. Using a custom optimization algorithm in MATLAB MPC toolbox is possible but not straightforward.

The NMPC and EMPC are approximately 3 to 6 times as fast as MATLAB counterparts. This is presumably explained by MATLAB's internal and the pendulum model. Indeed, many built-in MATLAB functions are not purely interpreted but contains calls to specialized compiled code. This is probably the case for \texttt{fmincon} solvers, the function used by MATLAB NMPC toolset. And, since the pendulum model is trivially simple, computing the predictions is not a performance bottleneck. Thus, in both languages, the majority of the time is spent in the optimization algorithms, both implemented in machine code for most parts. It is expected that the speed gap between the two languages would be larger for more complex plant models, since this part is mainly interpreted in MATLAB. Lastly, SLMPC is about 4 times faster than MATLAB. It is also 75 times faster than the nonlinear controllers on average: an impressive gain for similar closed-loop response!
% !TeX encoding = UTF-8
% !TeX spellcheck = en_US
\section{Conclusion}

This paper introduced a free and open-source package for advanced process control design in Julia. It relies on \texttt{ControlSystems.jl} and \texttt{JuMP.jl}, two powerful frameworks for computer-aided control system design and mathematical optimization, respectively. The new state estimator and predictive controller types are designed to be easy to use, clear and modular. The main features of the package were described and illustrated with two case studies. A benchmark comparison with the equivalent MATLAB toolbox exposes its computational efficiency, without any code generation, and thus avoiding the two-language problem. 

The package is still under active development and contributions are welcome. Short-term developments will focus on adding the support of differential-algebraic equations, as well as other transcription methods for the moving horizon estimation.
\section*{Acknowledgments}
The authors would like to acknowledge all the developers involved in the control system, automatic differentiation and optimization frameworks of Julia. Their ease of use and intuitiveness  significantly accelerated the programming of this package. 

\bibliographystyle{elsarticle-harv}
\renewcommand{\bibfont}{\small}
\bibliography{mpcPackageJulia_bibfile} 

\end{document}